\documentclass{aa}
\usepackage{epsfig}
\usepackage{graphics}
\usepackage{subfigure}
\usepackage{natbib}
\usepackage{amsmath}
\usepackage{enumerate}

\begin{document}

\title{Nonthermal processes and neutrino emission from the black hole GRO J0422+32 in a bursting state}

\author{F. L. Vieyro\inst{1,2}, Y. Sestayo\inst{3}, G. E. Romero\inst{1,2} \and J. M. Paredes\inst{3}}
  
\institute{Instituto Argentino de Radioastronom\'{\i}a (IAR, CCT La Plata, CONICET), C.C.5, (1984) Villa Elisa, Buenos Aires, Argentina \and Facultad de Ciencias Astron\'omicas y Geof\'{\i}sicas, Universidad Nacional de La Plata, Paseo del Bosque s/n, 1900, La Plata, Argentina \and Departament d'Astronomia i Meteorologia, Institut de Ci$\rm{\grave{e}}$ncies del Cosmos (ICC), Universitat de Barcelona (IEEC-UB), Mart\'{\i} i Franqu$\rm{\grave{e}}$s 1, E-08028 Barcelona, Spain}

\offprints{F. L. Vieyro \\ \email{fvieyro@iar-conicet.gov.ar}}

\titlerunning{Neutrino emission of GRO J0422+32}

\authorrunning{Vieyro, et al.}

\abstract
{GRO J0422+32 is a member of the class of low-mass X-ray binaries (LMXBs). It was discovered during an outburst in 1992. During the entire episode a persistent power-law spectral component extending up to $\sim 1$ MeV was observed, which suggests that nonthermal processes should have occurred in the system. }
{We study relativistic particle interactions and the neutrino production in the corona of GRO J0422+32, and explain the behavior of GRO J0422+32 during its recorded flaring phase. }
{We have developed a magnetized corona model to fit the spectrum of GRO J0422+32 during the low-hard state. We also estimate neutrino emission and study the detectability of neutrinos with 1 km$^3$ detectors, such as IceCube.}
{The short duration of the flares ($\sim$ hours) and an energy cutoff around a few TeV in the neutrino spectrum make neutrino detection difficult. There are, however, many factors that can enhance neutrino emission.}
{The northern-sky coverage and full duty cycle of IceCube make it possible to detect neutrino bursts from objects of this kind through time-dependent analysis.}

 
\maketitle

\section{Introduction}

The transient source GRO J0422+32 was discovered during an outburst by the Burst And Transient Source Experiment (BATSE, which covered the 20 - 1000 keV band) onboard the Compton Observatory in 1992 \citep{paciesas1992}. In only three days (August 5 to August 8), the intensity of the source rose from 0.2 Crab to 3 Crab in the $20-300$ keV range \citep{paciesas1992}. After the fast rise, the light curve showed an exponential decay on a timescale $\sim 40$ days. A secondary maximum $\sim 140$ days after the beginning of the outburst was reported by \citet{harmon1992}, which was also followed by an exponential decay. The entire episode lasted about 200 days (for a detailed description of the event see \citealt{ling2003}).

The outburst was also observed in UV/Optical/IR/Radio wavelengths (see, e.g., \citealt{castro-tirado1993,shrader1994,callanan1996,garcia1996}). The features of the optical and UV spectra show that GRO J0422+32 is a member of the class of low-mass X-ray binaries (LMXBs), see \citet{shrader1994}.  LMXBs usually undergo episodes where their X-ray luminosities increase in increments of up to several orders of magnitude. These episodes are thought to be produced by changes in the accretion rate onto a compact object. The outbursts typically last several months, as in the case of GRO J0422+32, although some can be as short as minutes, such as in V4641 Sgr, or as long as decades, as in GRS 1915+105, a source that has been active for the past 15 years and was quiescent before. Strong outbursts are recurrent, but LMXBs spend most of their lifetimes in quiescence \citep{shrader1994}. Many of these sources do not go through a soft X-ray phase \citep{brocksopp2004} and the outbursts represent transitions from the quiescent to the hard X-ray state \citep{esin1998}.

Including of GRO J0422+32 in the category of LMXBs is also supported by the detection of an optical counterpart by \citet{castro-tirado1993} during the outburst. The peak magnitude was $V \sim 13.2$, and subsequent measurements have shown that in quiescence the source dropped down to $V = 22.4$ \citep{zhao1994}.  Since the optical disk emission during outbursts overwhelms the light from the star \citep{sunyaev1993}, the orbital parameters are estimated in the quiescent state. The optical spectrum during quiescence shows the secondary to be a normal M0/M2V dwarf star \citep{filippenko1995,casares1995,beekman1997}.
  
The different measurements have yielded a wide range of estimates for the mass of the compact object. The mass function determined by several authors \citep{filippenko1995,casares1995,orosz1995} is even consistent with the presence of a neutron star. Assuming that the contamination of the accretion disk in the infrared band is negligible, \citet{callanan1996} estimated an orbital inclination of $i \leq 45^{\circ}$ and a mass of $\geq 3.4 M_{\odot}$ for the compact object. \citet{beekman1997}, however, suggested that the earlier estimates were fuzzed by the contribution of the disk, and they estimated a lower limit of $\sim 9 M_\odot$ for the compact object. This value strongly suggests there is a black hole in the system. More recently, \citet{reynolds2007} have detected some flickering from the accretion disk and, when including these effects, obtained a mass $\geq 10 M_\odot$, which strongly supports the black hole nature of the compact object.

The black hole nature of GRO J0422+32 is also supported by its spectrum. The X and $\gamma$-ray spectra measured by several instruments can be characterized by an exponentially truncated power law, with a photon power law index of $1.49\pm0.01$ and a cut-off energy of $132\pm2$ keV \citep{sunyaev1993,grove1998}. This spectrum is similar to the low-hard state spectrum of Cygnus X-1 \citep{poutanen1997,McConnell2000}, which is the most studied stellar black hole in our Galaxy. 

The X-ray spectra of galactic black holes, such as GRO J0422+32 and Cygnus X-1, have been the subject of several studies. The disk-corona model, in which soft photons emitted by the disk gain energy by successive Compton upscatterings in a hot corona, has successfully explained the spectra below $\sim1$ MeV (e.g., \citealt{poutanenSvensson96,narayan1994}). In addition, both GRO J0422+32 \citep{ling2003} and Cygnus X-1 \citep{McConnell2000} spectra show a persistent power law emission extended to $\sim1$ MeV. This fact suggests that nonthermal processes must be taking place in these systems, as was studied by \citet{li1996}, \citet{li1997}, and more recently by \citet{belmont2008}, \citet{malzac2009}, \citet{vurm2009}, \citet{flor01}, and \citet{vieyro2012}. 

Several attempts have been made to detect high-energy gamma-rays and neutrinos from XRBs, which are unambiguous evidence of the presence of a population of very energetic particles. Gamma-ray emission from confirmed accreting black holes has been detected only during transient events of two of the most famous high-mass XRBs: Cygnus X-3, detected up to $\sim 100$ GeV by the \textsl{AGILE} satellite \citep{tavani2009} and by Fermi/LAT \citep{abdo2009}, and an excess with a significance of 4.1$\sigma$ after trial correction of Cygnus X-1 up to $\sim$1 TeV in coincidence with an X-ray flare during its hard state \citep{albert2007}.  The intense radiation field of the companion star in both cases renders it difficult to detect gamma-rays above a few hundred GeV, owing to photon absorption in the stellar field. Neutrinos, on the other hand, could be the indication of very high-energy phenomena taking place in these objects, since they easily escape from the binary system. Several works have been devoted to studying the neutrino production in high-energy sources (e.g., \citealt{christiansen2006,kappes2007}), and current upper limits are at the level of $\sim 8 \times 10^{-11}$ TeV$^{-1}$cm$^{-2}$s$^{-1}$ for sources with soft spectra or with an energy cutoff below the PeV scale \citep{odrowski2011}. The recent completion of IceCube presents a unique opportunity to test our models.

In the case of LMXBs, absorption of gamma-rays is absent, so these are particularly suitable objects for studying both the gamma-ray and neutrino production. In this work we apply a hybrid strongly magnetized plasma corona model \citep{vieyro2012} to study the nonthermal processes and neutrino production during transient events of LMXBs. We use the source GRO J0422+32 as a prototype. Section \ref{analysis} summarizes the basic properties of the source in the quiescent state and during the outburst. In Sect. \ref{scenario} we briefly describe of the model. Then, we study first the source in the steady state (Sect. \ref{injection}), and in Sect. \ref{flare} we extend our model to encompass the possible enhanced emission of neutrinos and gamma-rays during a flare. Finally, in Sect. \ref{icecube}, we study the detectability of neutrinos from the source by making use of the latest performance studies of the IceCube neutrino detector \citep{odrowski2012}. Section 8 has the discussion.

\section{The binary system GRO J0422+32}\label{analysis}

The system GRO J0422+32 is a binary system with a low-mass star and a possible black hole. It has an orbital period of $5.1 \pm 0.01$ hours \citep{filippenko1995,callanan1996}. There is evidence that the system has a low inclination, $\sim 10^\circ  - 30^\circ$ \citep{beekman1997}. Based on a method that only depends on the spectral type of the companion star, \citet{esin1998} estimated a distance to the source of 2.6 kpc, which is consistent with previous results \citep{shrader1994}. 

Using this estimate of distance and the X-ray absorption \citep{filippenko1995}, the derived luminosity in the $0.5 –- 10.0$ keV band is $7.6\times 10^{30}$ erg s$^{-1}$ for the quiescent state \citep{garcia2001}. As expected for X-ray transients, the X-ray luminosity is several orders of magnitude higher during the outburst; in the $2-300$ keV, the luminosity reaches $5 \times 10^{37}$ erg s$^{-1}$ \citep{sunyaev1993}.

The X-ray spectrum of the source shows no evidence of an ultrasoft component. The column density is $N_{\rm{H}} < 2 \times 10^{21}$ cm$^2$ \citep{callanan1996}, which is too low a value to explain the absence of a soft component in the X-ray spectrum by absorption. This might indicate a peculiar orientation of the surface of the accretion disk with respect to the observer \citep{sunyaev1993}. 

During the outburst a flat radio emission was detected with the Very Large Array \citep{shrader1994}, which is a signature of an expanding environment. These observations, however, did not reveal any jet-like structure. More recently, the new capabilities of the Expanded Very Large Array were used to reach noise levels as low as 2.6 $\mu$Jy beam$^{-1}$ and to study black hole X-ray binaries in the hard and quiescent states \citep{MillerJones2011}. GRO J0422+32 was not detected to a $3\sigma$ upper limit of 8.3 $\mu$Jy beam$^{-1}$. The lack of clear radio emission might indicate the absence of a relativistic jet, leaving the corona as the most likely region for producing the nonthermal high-energy emission observed in the source.

\section{Basics of the model}

\subsection{Scenario}\label{scenario}

\begin{figure*}
\centering
\includegraphics[clip,width=0.6\textwidth, keepaspectratio]{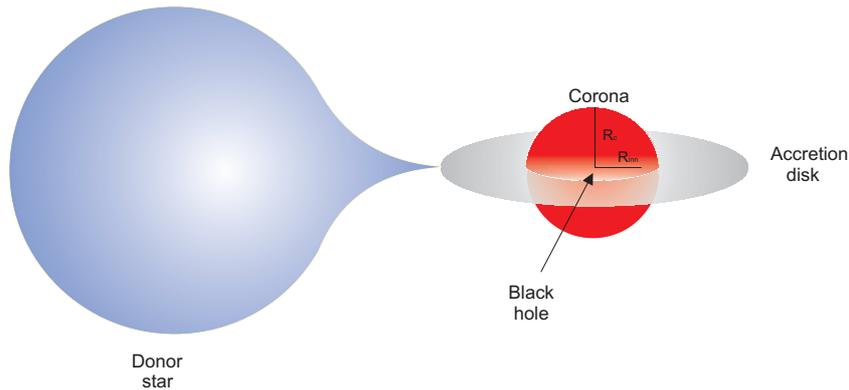}
\caption{Schematic representation of the components of the system discussed in the text. In the spherical corona, the thermal and nonthermal components are co spatial. Not to scale.}
\label{fig:geometria}
\end{figure*}

There are many similarities between the X-ray spectrum of GRO J0422+32 and Cygnus X-1\footnote{Cygnus X-1 has a well-resolved jet that is responsible for the radio emission \citep{stirling2001}. In this work, however, we are concerned with the emission at higher energies.}. Then, it is reasonable to assume as a first approximation that the mechanism responsible for the hard X-ray/soft $\gamma$-ray emission might be the same in both systems \citep{ling2003}. To explain the nonthermal power law spectrum of GRO J0422+32 in the active state, we propose a model of a magnetized black hole corona \citep{vieyro2012}. Figure \ref{fig:geometria} shows a sketch of the main components of the system. 

We assume a spherical corona with a radius $R_{\rm{c}}$ and an accretion disk that penetrates the corona up to $R_{\rm{inn}}<R_{\rm{c}}$. In corona+disk models, the size of the emitting region can be constrained by considering some features of the spectrum. The transition radius between the hot corona and the soft photons emission region is estimated to be $\sim 200 R_{\rm{g}}$ \citep{esin1998}, where $R_{\rm{g}}$ is the gravitational radius ($R_{\rm{g}}=GM/c^2$). We assume an inner radius for the disk of $R_{\rm{inn}} = 200 R_{\rm{g}}$. Typically the ratio between the inner radius of the disk and the corona radius is 0.9 \citep{poutanen1998}. Here, we adopt a corona radius of $R_{\rm{c}} = 220 R_{\rm{g}}$. We suppose that the corona is homogeneous and isotropic. First, we study the corona in a low-hard steady state, and in Sect. \ref{flare} we extend our model to the corona during the flare.

Since there is no information regarding the emission of the disk, we use the disk model for the low-mass binary XTE J1118+480 developed by \citet{vila2012}. The corresponding temperature of the accretion disk is $ \sim 0.08$ keV. In a recent study of the ultraviolet spectra of galactic black holes in quiescence \citep{hynes2012}, the estimated temperature of the accretion disk of GRO J0422+32 is considerably lower than the value adopted in our work. This result, however, does not disagree with our assumption. According to ADAF models of black hole coronae, the value of $R_{\rm{inn}}$ is lower in the low-hard state than the one in quiescence, so that the temperatures achieved by the accretion disk are higher and similar to the one adopted here for the low-hard state.

The hard X-ray emission of the corona is characterized by a power law with an exponential cutoff at high energies:

			\begin{equation}
				n_{\rm{ph}}(E)=A_{\rm{ph}} E^{-\alpha}  e^{-E/E_{\rm{c}}}\; \textrm{erg$^{-1}$ cm$^{-3}$.}
			\end{equation}

\begin{figure}
\centering
\includegraphics[clip,width=0.4\textwidth, keepaspectratio]{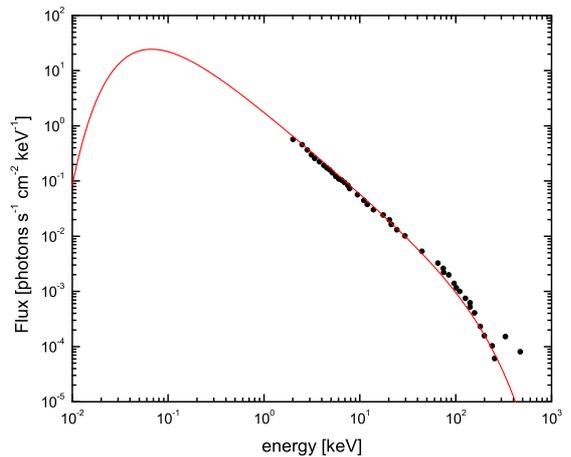}
\caption{The spectrum of GRO J0422+32 observed with instruments onboard the \textit{Mir-Kvant} observatory, fitted with a power law of index $1.49$ and an exponential cut-off 132 keV \citep{sunyaev1993}.}
\label{fig:thermal}
\end{figure}

\noindent Using the data taken with the Mir-Kvant observatory during the 1992 outburst, we fit the spectrum with a photon index of $\sim 1.49$ and a cut-off energy $E_{\rm{c}}=132$ keV, as shown in Fig. \ref{fig:thermal} \citep{sunyaev1993,grove1998}. The normalization constant $A_{\rm{ph}}$ can be obtained from $L_{\rm{c}}$; at a distance of $2.6$ kpc, the corona luminosity is $\sim 5 \times 10^{37}$ erg s$^{-1}$ \citep{sunyaev1993}. This value is equivalent to 5 \% of the Eddington luminosity of a $\sim 9M_{\odot}$ black hole, which is the value for the mass adopted in our model.

We assume that the corona consists of a two-temperature plasma, with an electron temperature $T_{e}=10^{9}$ K and an ion temperature $T_{i}=10^{12}$ K \citep{narayan1995a,narayan1995b}. The values of the main parameters are obtained by assuming equipartition of energy between the magnetic field, the bolometric luminosity of the corona and the kinetic energy of the plasma, as described in \citet{vieyro2012}. Table \ref{table} summarizes the values of the different parameters derived or assumed for GRO J0422+32.

\begin{table}
    \caption[]{Main parameters of the model.}
   	\label{table}
   	\centering
\begin{tabular}{ll}
\hline\hline 
Assumed parameters & Value\\ [0.01cm]
\hline
$M_{\rm{BH}}$:  black hole mass [$M_{\odot}$]							& $9$					        \\
$R_{\rm{c}}$:   corona radius [cm] 										 	  & $220 R_{\rm{g}}$   	\\
$T_{e}$:        electron temperature [K] 							  	&	$10^9$\tablefootmark{\star}			\\
$T_{i}$:        ion temperature [K] 											&	$10^{12}$\tablefootmark{\star}		\\
$E_{\rm{c}}$:   X-ray spectrum cutoff [keV]							  & $132$       				\\
$\alpha$: 			X-ray spectrum power law index    				& $1.49$							\\
$kT$:						disk characteristic temperature [keV] 	  & $0.08$\tablefootmark{\star \star}			\\
$n_{i},n{e}$:   plasma density [cm$^{-3}$] 								& $\sim 10^{12}$    	\\
$B_{\rm{c}}$: 	magnetic field [G]				 								& $\sim 10^5$	  			\\
$v_{\rm{rec}}$: reconnection speed [c]                    & $0.5$ 							\\
$\eta$: 				acceleration efficiency 									& $0.08$			  			\\
\hline\hline
Free parameters & Value\\ [0.01cm]
\hline
$q$: 						fraction of power injected in relativistic particles 	& $0.12$		    				\\
$a$: 						hadron-to-lepton energy ratio 						& $100$		    				\\


\hline  \\[0.005cm]
\end{tabular}	
\tablefoot{
\tablefoottext{\star}{Typical value for ADAF corona model \citep{narayan1995a,narayan1995b}.}\\
\tablefoottext{\star \star}{Typical value for LMXBs accretion disks \citep{vila2012}}.
 } 	 																				
\end{table}

\subsection{Particle acceleration and losses}\label{losses}

We now consider the interaction of locally injected relativistic particles with the matter, photons, and magnetic fields of the corona and the disk, which are taken as background components. The relevant processes of interaction of relativistic particle are the following: interaction with the magnetic field producing synchrotron radiation; collisions with the corona plasma via relativistic Bremsstrahlung, for electrons and muons, and via hadronic inelastic collisions for protons and charged pions; and interactions with the photon fields through IC scattering (electrons and muons) and photomeson production (protons and charged pions). The photon fields considered initially as targets are the X-ray photon field of the corona and the field produced by the accretion disk; then, the nonthermal photons are added to the target field. The stellar field is not an efficient target \citep{vieyro2012}.

We also consider the injection of electron/positron pairs through three different channels: photopair production, Bethe-Heitler process, and muon decay. Pairs can also annihilate. Convenient expressions for the energy loss rates for all these processes can be found, e.g., in \citet{vieyro2012} and \citet{vilaaha}.

In \citet{flor01}, it is shown that in a corona dominated by advection most hadrons fall onto the black hole before cooling, thus the radiated luminosity is not sufficient to explain the nonthermal tail detected in galactic black holes. Then, we consider a static corona supported by magnetic fields (see \citealt{beloborodov1999}), where the relativistic particles can be removed by diffusion. In particular, we refer to static corona as those systems where the advection time scale is $ > 10$ s. This is the value expected in highly magnetized systems, much greater than those associated with advective corona where the mean radial velocity is the free-fall velocity, $v \sim 0.1c$, and the advection time scale is on the order of 1 s \citep{flor01}. 

For a static corona, the diffusion rate is

	\begin{equation}\label{eq:diff}
				t_{\rm{diff}}^{-1}=\frac{2D(E)}{R_{\rm{c}}^2} ,
		\end{equation}
		
\noindent where	$D(E)$ is the diffusion coefficient	given by $D(E)=r_{\rm{g}}c/3$ in the Bohm regime, and $r_{\rm{g}}=E/(eB)$ is the giroradius of the particle.
	
 The maximum energy that a relativistic particle can attain depends on the acceleration mechanism and the different processes of energy loss. The acceleration rate $t^{-1}_{\rm{acc}}=E^{-1}dE/dt$ for a particle of energy $E$ in a magnetic field $B$, is given by

\begin{equation}
	t^{-1}_{\rm{acc}}=\frac{\eta ecB}{E},
	\label{accrate}
\end{equation}
   
\noindent where $\eta\leq 1$ is a parameter that characterizes the efficiency of the acceleration. According to a model of magnetic reconnection as an acceleration mechanism, the acceleration efficiency can be obtained by \citep{delValle2011}

\begin{equation}
\eta \sim 0.1 \frac{r_{g}c}{D} \Big( \frac{v_{\rm{rec}}}{c} \Big)^2,
\end{equation}

\noindent where $v_{\rm{rec}}$ is the reconnection speed, which has a value similar to the Alfv\'en speed in violent reconnection events. In a corona characterized with the parameters of Table \ref{table}, the Alfv\'en speed is $\sim 0.5c$ \citep{bette2005}, yielding an acceleration efficiency of $\eta \sim 0.08$. We note that the acceleration mechanism is not included as an energy loss term in the transport equation, so it is only used to determine the injection function of primary electrons and protons.
			
\begin{figure*}[!ht]
\centering
\subfigure[Electron losses.]{\label{fig:perdidas:a}\includegraphics[width=0.45\textwidth,keepaspectratio]{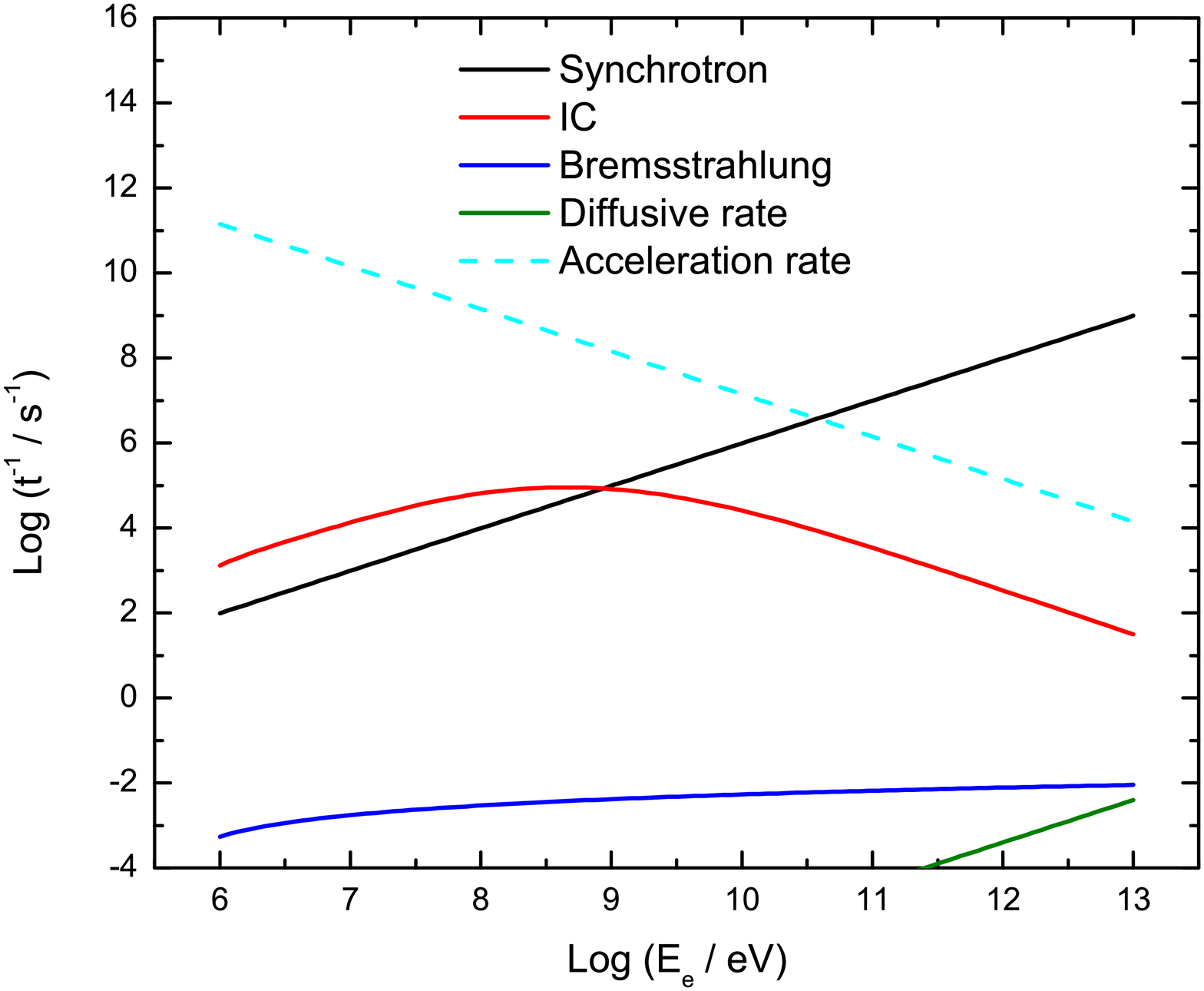}} \hspace{20pt} 
\subfigure[Proton losses.]{\label{fig:perdidas:b}\includegraphics[width=0.45\textwidth,keepaspectratio]{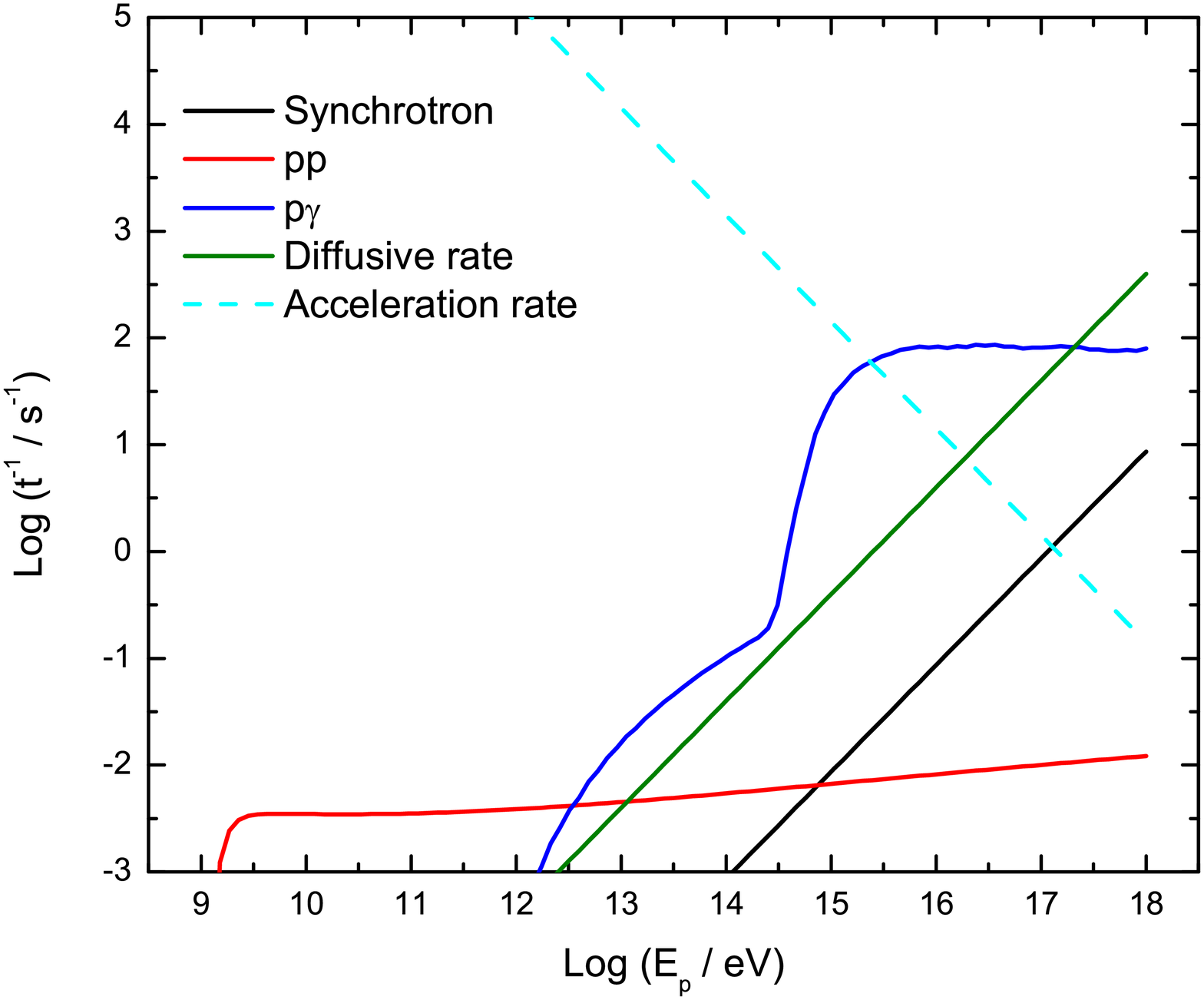}} \hfill \\ 
\subfigure[Pion losses.]{\label{fig:perdidas:c}\includegraphics[width=0.45\textwidth, keepaspectratio]{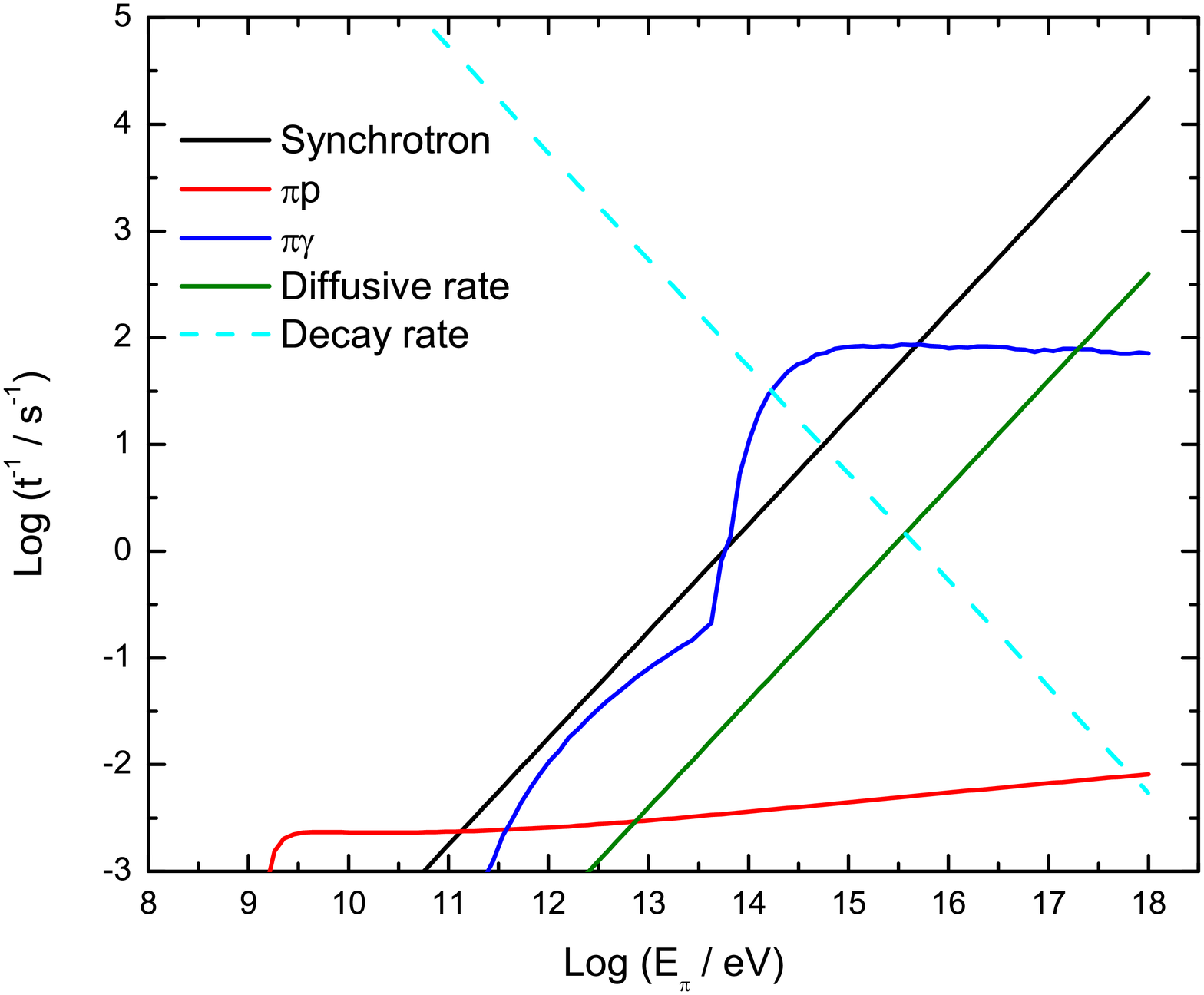}}  \hspace{20pt}
\subfigure[Muon losses.]{\label{fig:perdidas:d}\includegraphics[width=0.45\textwidth, keepaspectratio]{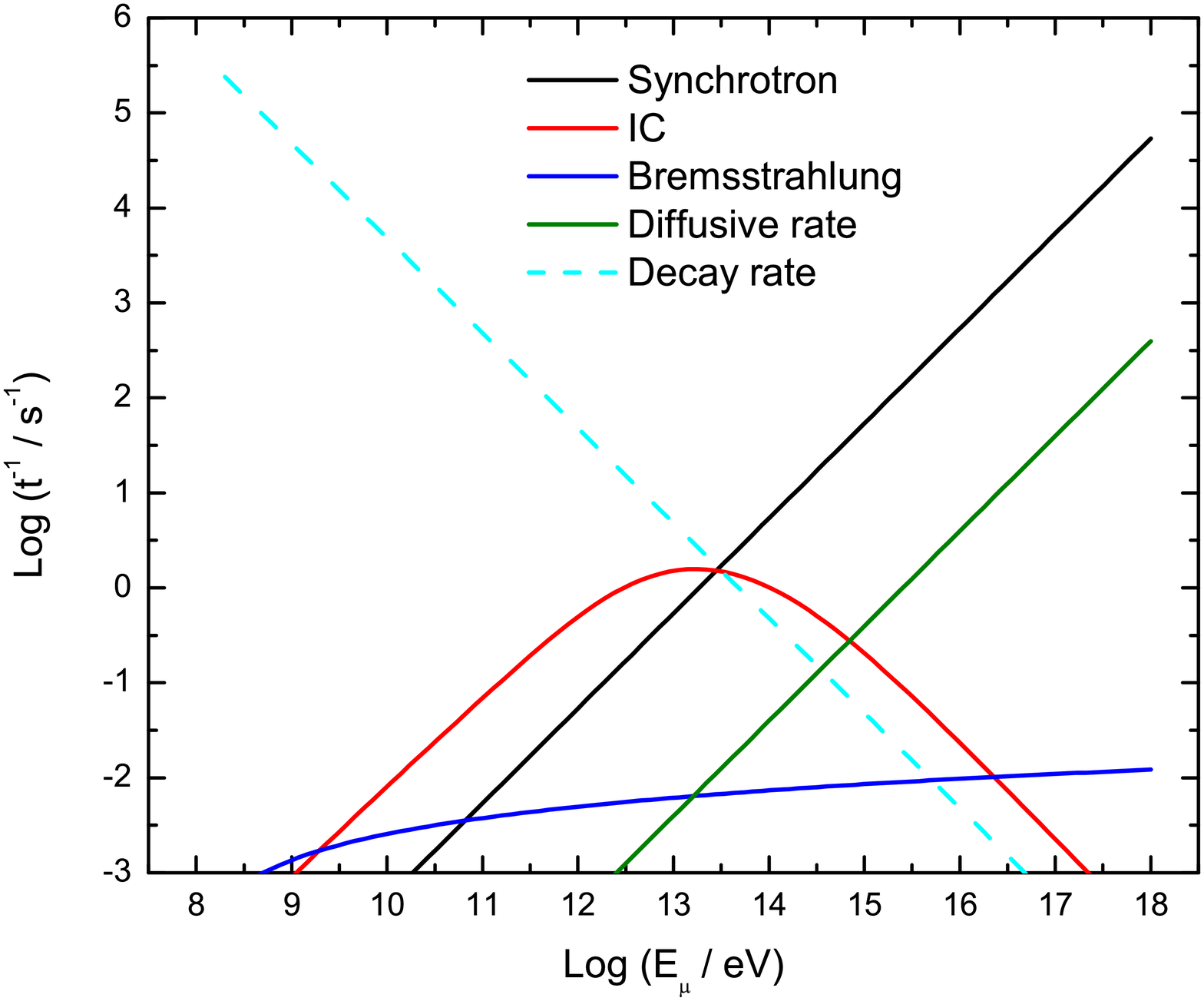}} \hfill
\caption{Energy losses in a corona characterized by the parameters of Table \ref{table}.}
\label{fig:perdidas}
\end{figure*}

	Figure \ref{fig:perdidas} shows the cooling rates for different energy-loss processes, together with the acceleration and escape rates, for each type of particle considered. 
	
Under the physical conditions previously described, the main channels of energy loss for electrons are IC scattering and synchrotron radiation. For protons, both $pp$ and $p \gamma$ interactions are relevant. Charged pions decay before cooling, which leads to the production of energetic muons and neutrinos. The diffusion has almost no effect on the various particle distributions.

The maximum energies obtained by electrons and protons are $E_{\rm{max}}^{(e)}\approx 40$ GeV and $E_{\rm{max}}^{(p)}\approx 3.9\times 10^{15}$ eV, respectively. These values are compatible with the Hillas criterion, given the size of the corona \citep{hillas1984}. With these values, high-energy protons may lead to the production of neutrinos of energies above 0.1 TeV, which is the current IceCube detection threshold.

\section{Spectral energy distributions}\label{injection}

In \citet{ling2003} a detailed analysis of the outburst of GRO J0422+32 is presented, showing that the spectrum of the source changed its shape along the episode. At the beginning, the spectrum can be described by a simple power law of index $\alpha \sim 1.75$ (flux $\propto E^{-\alpha}$), and then it clearly shows two components, a thermal component plus a high energy power law with a variable index \citep{ling2003}.

\begin{figure*}[!ht]
\centering
\subfigure[]{\includegraphics[width=0.45\textwidth,keepaspectratio]{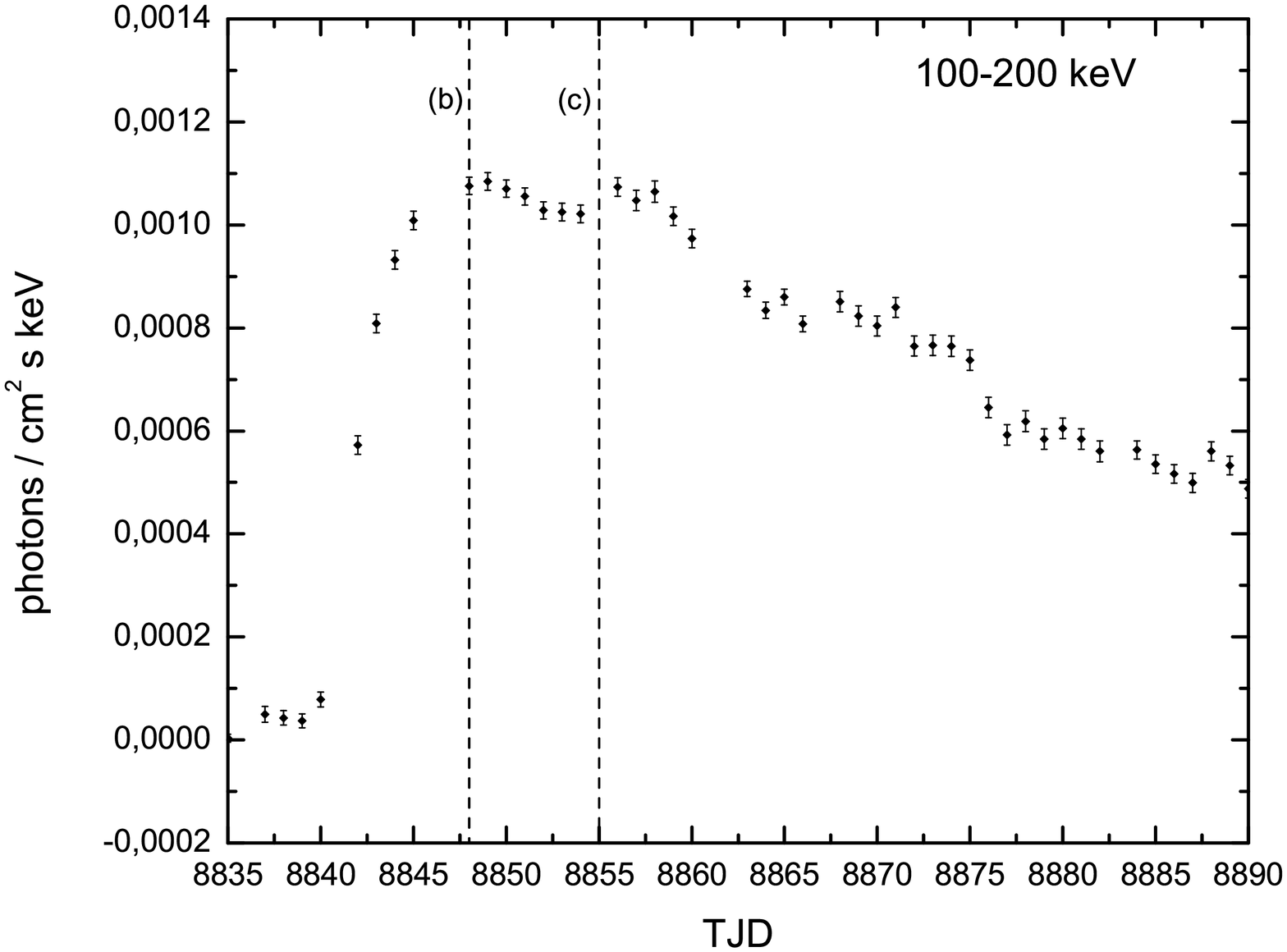}} \hspace{20pt} 
\subfigure[]{\includegraphics[width=0.45\textwidth,keepaspectratio]{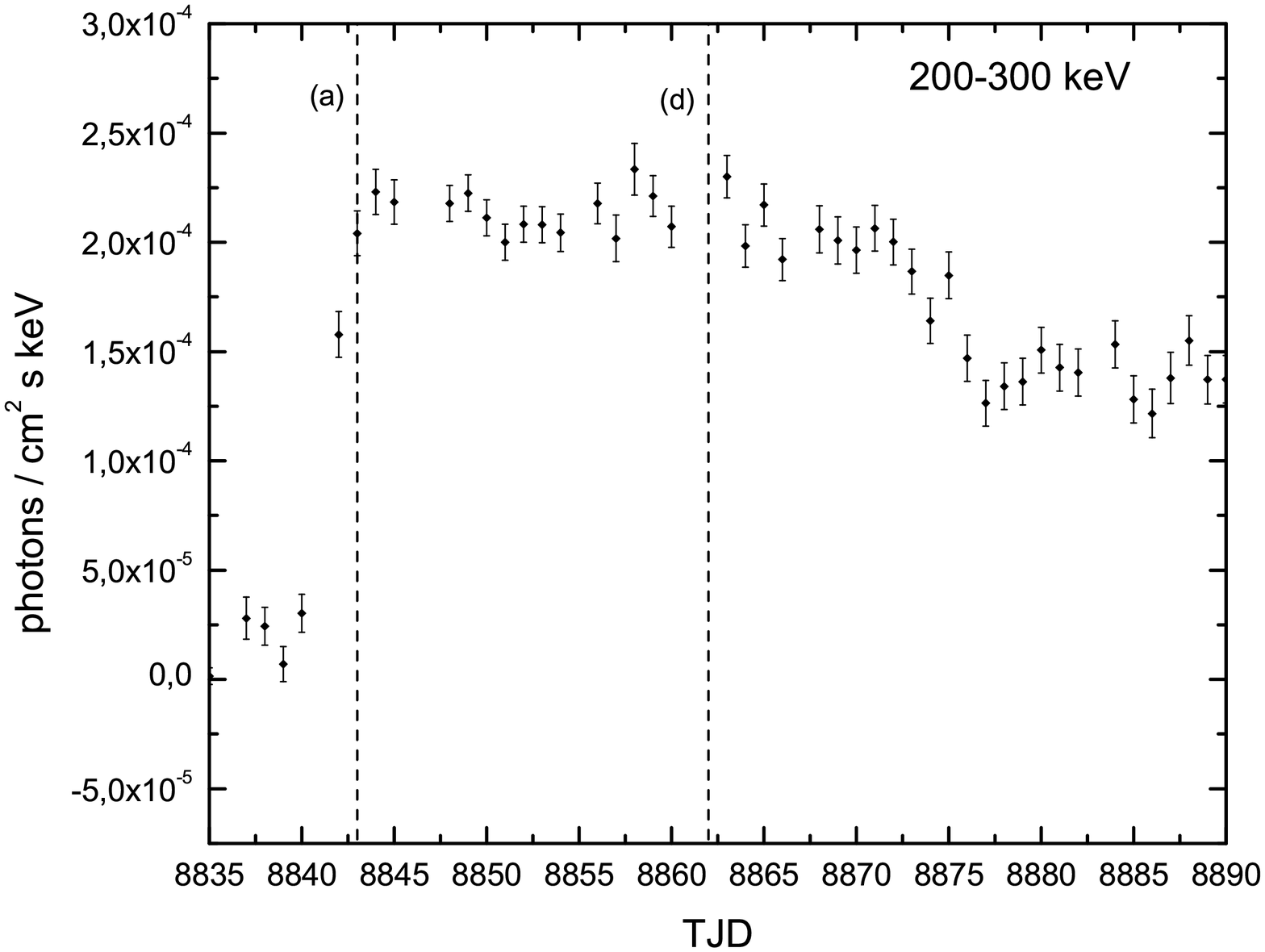}} \hfill \\ 
\caption{Flux histories detected with BATSE in the energy bands $100-200$ keV (left panel) and $200-300$ keV (right panel). The data is from \citet{ling2003}. The plateau phase is the period of $\sim 15$ days between TJD $\sim 8845$ and TJD $\sim 8860$. Lines (b) and (c) correspond to TJD 8848 and 8855, respectively, where the two first maxima at $E <200$ keV were detected; in an analogous way, lines (a) and (d) correspond to TJD 8843 and 8862, where the two first maxima at $E > 200$ keV were detected.}
\label{fig:fluxes}
\end{figure*}

The variations in the spectrum might reflect changes in the injection function of the relativistic particles. We applied our model to reproduce the spectrum during the plateau phase of the event, associated with the low-hard state. The plateau phase can be seen in Fig. \ref{fig:fluxes} around the peak of the luminosity (TJD $\sim 8850$). We selected the observations made with BATSE on TJD 8843 to fit the spectrum, given that these are the data with smaller observational errors.  

\citet{ling2003} also detected time- and energy-dependent flux variability, and obtained different temporal features for different energy bands. For example, in Fig. \ref{fig:fluxes} lines (b) and (c) show the first two maxima in the energy band $35-200$ keV, whereas lines (a) and (d) indicate the first two maxima detected at energies above 200 keV.

To obtain the particle distributions and to compute the spectral energy distribution (SED), we implement a consistent treatment of nonthermal emission in the magnetized corona \citep{vieyro2012}. This method solves the set of coupled differential equations for all kinds of particles through an iterative scheme.

The injection function for nonthermal protons and electrons is a power law of the energy of the particles given by

\begin{equation}\label{eq:injection}
Q(E)=Q_{0} E^{-\alpha}e^{-E/E_{\rm{max}}},
\end{equation}

\noindent as a consequence of multiple, fast, magnetic reconnection events. Following \citet{bette2005}, we  adopt a standard index $\alpha=2.2$ (see, nonetheless, \citealt{drury2012} for critics and alternatives). The normalization constant $Q_{0}$ can be obtained from the total power injected in relativistic protons and electrons, $L_{\rm{rel}}=L_{p}+L_{e}$. The power injected in relativistic particles is considered to be a fraction of the corona luminosity $L_{\rm{rel}} = q L_{\rm{c}}$.  

The total power available to accelerate particles through magnetic reconnection in a magnetized system  can be estimated as in \citet{delValle2011},

\begin{equation}
L = \frac{B^2}{8\pi } A v_{\rm{A}},
\end{equation}

\noindent where $A \sim 4\pi R_{\rm{c}}^2$, and $v_{\rm{A}}$ is the Alfv\'en velocity. In our model $L$ is $\sim 15$ \% $L_{\rm{c}}$. The precise way in which power is divided between hadrons and leptons is unknown, but different scenarios can be taken into account by setting $L_{p}=aL_{e}$. We consider a model with $a=100$ (proton-dominated scenario, as for Galactic cosmic rays), where the high hadron content favors the neutrino production.

\begin{figure}
\centering
\includegraphics[clip,width=0.5\textwidth, keepaspectratio]{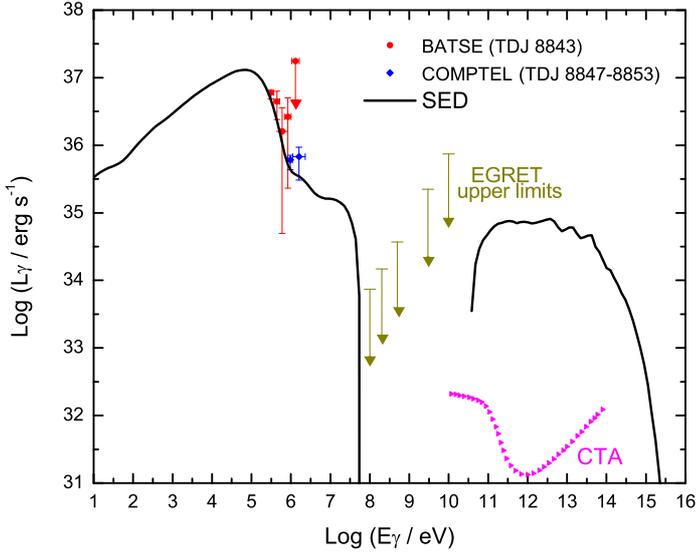}
\caption{Spectrum of GRO J0422+32 detected with BATSE \citep{ling2003} and fitted with the spectral energy distribution obtained with a nonthermal corona model. The values of the free parameters in this adjustment are $a=100$ and $q=0.12$.}
\label{fig:steadyState}
\end{figure}

In Fig. \ref{fig:steadyState} we show the data of GRO J0422+32 in TJD 8843 fitted with the final SED obtained with our model. The best-fit value of the parameter $q$ is 0.12; that is 12 \% of the total power available to accelerate particles through magnetic reconnection is injected in relativistic particles.

\begin{figure}
\centering
\includegraphics[clip,width=0.5\textwidth, keepaspectratio]{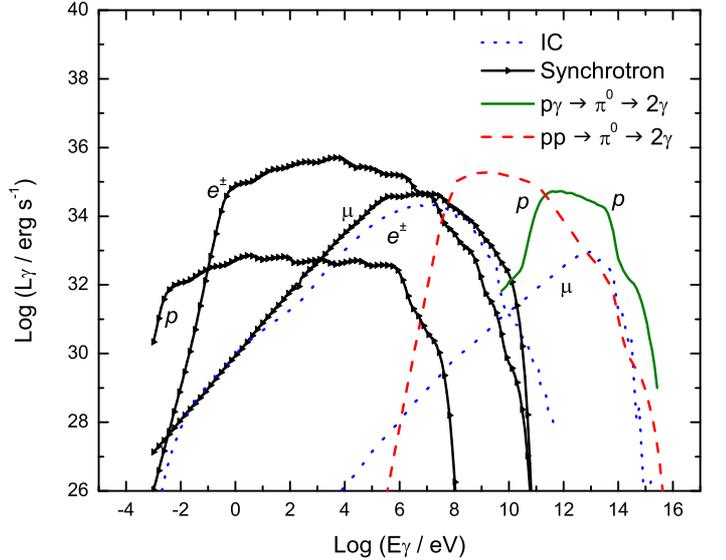}
\caption{Main contribution to the nonthermal luminosity. Internal absorption of the radiation is not included, but it has a strong impact on the final SED as shown in the previous figure.}
\label{fig:mainSEDs}
\end{figure}

The synchrotron radiation of electron/positron pairs and protons dominates the spectrum at low energies; at energies $\sim 10^{6-8}$ eV, IC scattering of electron/positron pairs is also relevant. At high energies the main contribution is the decay of neutral pions produced in hadronic interactions (both $pp$ and $p\gamma$).

In our model the internal absorption suppresses the emission completely for energies in the band $10^8 < E < 3\times10^{10}$ eV. Unlike a jet-model, which yields emission in this energy range \citep{vila2008}, our result agrees with the non detection of GRO J0422+32 by EGRET (which was the instrument operating when the episode occurred, and covered the energy range $\sim$100 MeV $< E < 30$ GeV). The values of the EGRET upper limits for GRO J0422+32 are from \citet{levinson1996}.

The luminosity of an M2V dwarf star during quiescence is $\sim 10^{32}$ erg s$^{-1}$ and the effective temperature is $3500$ K \citep{reid1995}. During the outburst, the optical magnitude of the companion star increased in 9 mag, which corresponds to a change in the luminosity $>3$ orders of magnitude. The maximum luminosity was $6 \times 10^{35}$ erg s$^{-1}$, and the peak of the emission appeared at $0.15$ eV. On the one hand, the optical emission of the star during the outbursts was negligible in comparison with the disk emission \citep{sunyaev1993}, and on the other, in order to create electron/positron pairs in the stellar field, the energy of the gamma photon should be $ > 10^{15}$ eV, which is very unlikely according to our model. Then, the absorption in the stellar field was not relevant.

\section{Flare}\label{flare}

It is known that X-ray binaries undergo transient radiative flares (see, e.g., \citealt{stern2001,mazets1996,golenetskii2003,albert2007,tavani2009} for flare events in Cygnus X-1). In particular, within the first 80 days of the 1992 outburst of GRO J0422+32, four shorter but strong episodes were detected in the energy band $0.4-1$ MeV (\citealt{ling2003}). In Fig. \ref{fig:fluxes} lines (a) and (b) indicate the first two of these episodes. These are the flares that took place during the plateau phase of the main episode. The other two flares occurred when the corona luminosity decreased to half the peak level and 1/10 of the peak (i.e. at $\sim 30$ and $\sim 80$ days after the beginning of the outburst, respectively). Their contribution to neutrino emission is then negligible in comparison with the flares occurring in the plateau. 

A possible cause of these events might be an increase in the power injected in relativistic particles, owing to large-scale reconnection events. This suggestion is supported by observations of solar flares where magnetic reconnection can trigger diffusive acceleration \citep{tsuneta1998,lin2008,kowal2011}. To represent a sudden injection of relativistic particles, we adopt the following analytic expression \citep{reynoso2010}: 

\begin{align}
Q(E,t) = &  Q(E)(1-e^{t/\tau_{\rm{rise}}} ) \nonumber \\
& \times \left[ \frac{\pi}{2}- \arctan \Big( \frac{t-\tau_{\rm{plat}}}{\tau_{\rm{dec}}} \Big) \right],
\end{align}

\noindent where $\tau_{\rm{rise}}$, $\tau_{\rm{dec}}$ and $\tau_{\rm{plat}}$ are the rise, decay, and plateau duration, respectively. Since BATSE spectra are daily, we consider flares of a duration of less than a day; we adopt $\tau_{\rm{rise}} = 30$ min, $\tau_{\rm{dec}} = 1$ h, and $\tau_{\rm{plat}} = 2$ h for a rapid flare. 

The energy dependence is the same as in the steady state, given by Eq. \ref{eq:injection}. The normalization constant $Q_{0}$ is again obtained from the total power injected in relativistic protons and electrons. It is assumed that the thermal corona remains unaffected during the event. In our model, the power injected in the flare is 15\% of the luminosity of the corona, which is equal to the total power available for accelerating particles via reconnection events. Such energetic flares are known from studies of solar flares \citep{lin2008}.

\begin{figure}
\centering
\includegraphics[clip,width=0.5\textwidth, keepaspectratio]{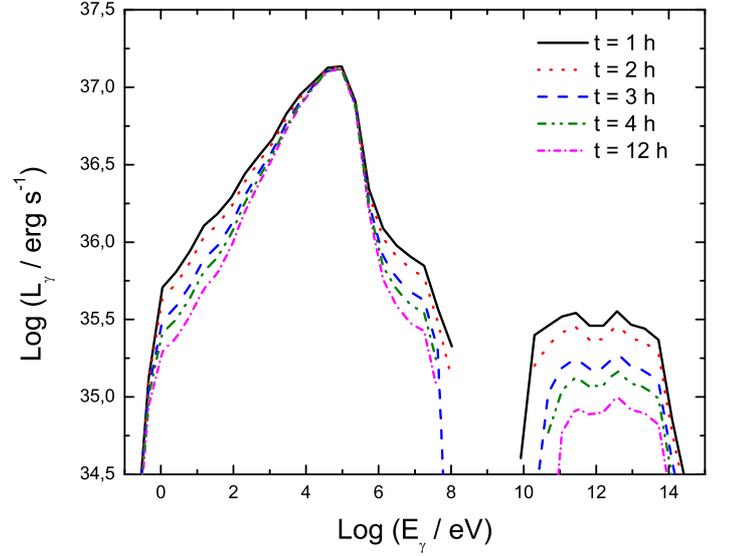}
\caption{Evolution of the luminosity during a flare of $\sim 2$ h of duration. Since the cooling time scales in the coronae are significantly shorter than flare time scales, the shape of the spectrum does not change and just shows decreasing luminosity levels as the flare evolves. 
}
\label{fig:flareLuminosity}
\end{figure}

In Fig. \ref{fig:flareLuminosity} we show the evolution of the luminosity of the source along a span of 12 hours. That the shape of the spectrum remains almost unaltered along the outburst can be explained as follows: cooling time scales in galactic black hole coronae are significantly shorter than flare time scales (typically of hours or even days, see e.g. \citealt{malzac2000}). Then, the system behaves as in the steady state, but shows decreasing luminosity levels as the flare evolves. The most noticeable changes occur at high energies. As in the steady state, the high-energy emission is due to hadronic interactions, and since we are considering a proton-dominated corona ($a=100$), the increment in the power injected in relativistic particles (from 10 \% $L_{\rm{c}}$ in steady state to 15 \% $L_{\rm{c}}$ at the maximun of the outburst) can be directly seen as an increase in the luminosity ($\sim 0.5$ order of magnitude) at energies above 10 GeV.

\section{Neutrino emission}\label{neutrino}

We are interested in estimating the $\nu_{\mu}$ production, since the searches for point-like neutrino emission make use of this neutrino flavor. As explained in the following section, however, we also need to estimate the production of all neutrino flavors in order to take the effects of neutrino oscillations into account. Then, we consider $\nu_{e}$ production by the channel of muon decay

	\begin{equation}
		 \mu^{\pm} \rightarrow e^{\pm} + \overline{\nu}_{\mu}(\nu_{\mu}) + \nu_{e}(\overline{\nu}_{e}) ,
		\end{equation}
		
\noindent and $\nu_{\mu}$ production by the previous channel plus charged pion decay

\begin{equation}
		 \pi^{\pm} \rightarrow \mu^{\pm} + \nu_{\mu}(\overline{\nu}_{\mu}) \textrm{.}
		\end{equation}
		
\noindent Thus, the total emissivity of muon-neutrinos is \citep{reynoso2009}

\begin{equation}
\phi_{\nu_{\mu}}(E,t) = \phi_{\pi \rightarrow \nu_{\mu}}(E,t) + \phi_{\mu \rightarrow \nu_{\mu}}(E,t) ,
\end{equation}

\noindent where

	\begin{equation}\label{eq:pi_mu}
	\begin{aligned}
		\phi_{\pi \rightarrow \nu_{\mu}}(E,t) = \int^{E^{\rm{max}}}_{E/(1-r_{\pi})} & dE_{\pi} t^{-1}_{\pi,\rm{ dec}}(E_{\pi})N_{\pi}(E_{\pi},t) \\       
		 & \times \frac{1 }{E_{\pi}(1-r_{\pi})},
		 \end{aligned}
	\end{equation}
	
\noindent with $r_{\pi}=(m_{\mu}/m_{\pi})^2$ and

	\begin{align}
		\phi_{\mu \rightarrow \nu_{\mu}}(E,t) &= \sum^4_{i=1} \int^{E^{\rm{max}}}_{E} \frac{dE_{\mu}}{E_{\mu}} t^{-1}_{\mu,\rm{ dec}}(E_{\mu})N_{\mu_{i}}(E_{\mu},t)\\
		 &\times \left[ \frac{5}{3}-3x^2+ \frac{4}{3}x^3 \right]. \nonumber
	\end{align}
	
\noindent In this expression, $x=E/E_{\mu}$, $\mu_{\{1,2\}}=\mu_{\rm{L}}^{\{-,+\}}$, and $\mu_{\{3,4\}}=\mu_{\rm{R}}^{\{-,+\}}$.

In a similar way to Eq. \ref{eq:pi_mu}, the total emissivity of electron-neutrinos $\nu_{e}$ is \citep{lipari2007}:

	\begin{equation}
	\begin{aligned}
		\phi_{\mu \rightarrow \nu_{e}}(E,t) = \int^{E^{\rm{max}}}_{E} & dE_{\mu} t^{-1}_{\mu,\rm{dec}}(E_{\mu})N_{\mu}(E_{\mu},t) \\
		&\times \frac{ F_{\mu \rightarrow \nu_{e}} (E / E_{\mu} )}{E_{\mu}} ,
		\end{aligned}
	\end{equation}
	
\noindent where

	\begin{equation}
		F_{\mu \rightarrow \nu_{e}} (x) = 2-6x^2+4x^3,
	\end{equation}
	
\noindent for unpolarized muons.
	
The third flavor of neutrino, $\nu_{\tau}$, can be injected in astrophysical sources, for example, by a different channel of electron/positron annihilation:

	\begin{equation}
		 e^{+} + e^{-} \rightarrow \tau^{+} + \tau^{-}.
		\end{equation}

\noindent This channel, however, takes place only at energies of the electrons and positrons $ > 100$ GeV, but this channel does not occur in our model (see Fig. \ref{fig:perdidas}), so we consider a source with null production of initial $\nu_{\tau}$.

\subsection{Neutrino oscillations}

Neutrinos from astrophysical sources detected on Earth travel long distances, and the probability of neutrino oscillations depends on distance. Then the flux of neutrinos of a given flavor can be affected by this effect. If we call $\phi^{0}_{\alpha}$ to the neutrino flux of flavor $\alpha$ at the source, then the arriving flux on Earth is \citep{esmaili2010}

\begin{equation}
\phi_{\alpha} = \sum_{\beta = e, \mu, \tau} P_{\alpha \beta} \phi^{0}_{\beta},
\end{equation}

\noindent where $P_{\alpha \beta}$ is the oscillation probability, and for long distances to the source is given by

\begin{equation}
P_{\alpha \beta} = \sum_{j=1}^{3} |U_{\alpha j}|^2|U_{\beta j}|^2. 
\end{equation}

\noindent Here $U_{\alpha j}$ is the mixing matrix. The current best-fit parameters of the mixing matrix are listed in Table \ref{table2}.

\begin{table}[!t]
    \caption[]{Best fit for the mixing angles \citep{nakamura2010}.}
\begin{tabular}{ccc}
\hline\noalign{\smallskip}
Parameter & Best-fit & Current allowed range $(3\sigma)$          \\[0.01cm]
\hline\noalign{\smallskip}
$\sin^2(\theta_{12})$  & $0.304$  &  $0.25-0.37$					        \\[0.01cm]
$\sin^2(\theta_{23})$  & $0.500$  &  $0.36-0.67$					        \\[0.01cm]
$\sin^2(\theta_{13})$  & $0.014$  &  $\leq 0.056$					        \\[0.01cm]

\hline\\[0.005cm]
\end{tabular}	
  \label{table2}
\end{table} 

One aim of our work is to predict what might be detected by IceCube if the source GRO J0422+32 flared as it did it in 1992. Since the effect of neutrino oscillations can change the flux of $\nu_{\mu}$ estimated in Sect. \ref{neutrino}, we correct it by applying the analysis described above.

The final values of the mixing matrix are taken from \citet{vissani2011}, and the final neutrino flux results:

\begin{eqnarray}
\phi_{\nu_{\mu}} & = & P_{\mu e} \phi^0_{e} + P_{\mu \mu} \phi^0_{\mu} + P_{\mu \tau} \phi^0_{\tau} \\
           & = & 0.221 \phi^0_{e} + 0.390 \phi^0_{\mu} + 0.390 \phi^0_{\tau}.
\end{eqnarray}

\noindent Though the precise values of the parameters depend on the analysis, the changes in the mixing matrix are not very large \citep{vissani2011}.

The differential flux of neutrinos arriving at the Earth can be obtained as

\begin{equation}
\frac{d \Phi_{\nu_{\mu}}}{dE} = \frac{1}{4\pi d^2} \int_{\rm{V}}{d^3r \phi_{\mu}(E,t)} .
\end{equation}

\begin{figure}
\centering
\includegraphics[clip,width=0.5\textwidth, keepaspectratio]{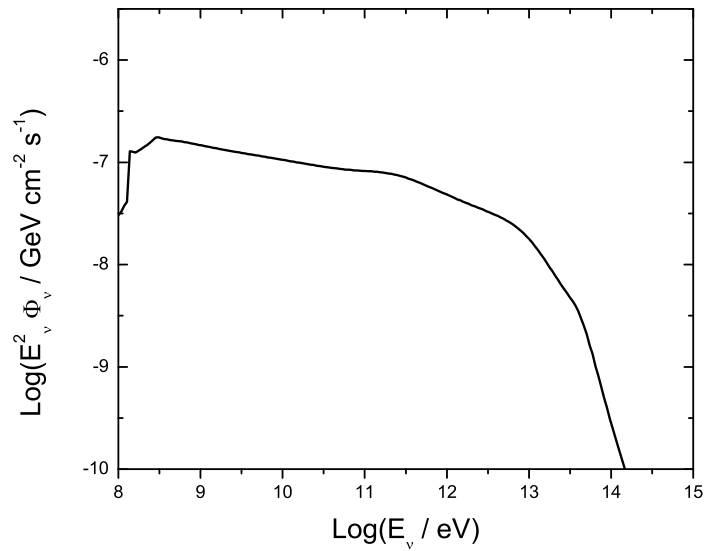}
\caption{Differential flux of muon neutrinos produced in the hard state of GRO J0422+32 arriving at the Earth. The effects of neutrino oscillations are included.}
\label{fig:neutrinoSteady}
\end{figure}

\begin{figure}
\centering
\includegraphics[clip,width=0.5\textwidth, keepaspectratio]{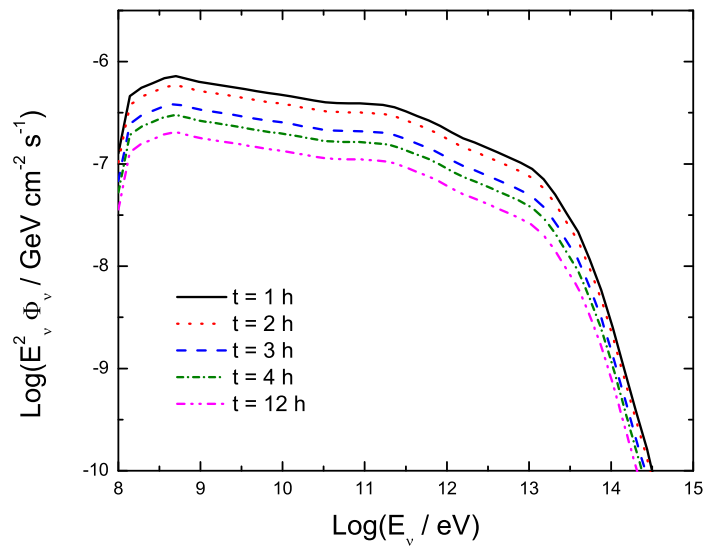}
\caption{Differential flux of muon neutrinos from a flare in GRO J0422+32 arriving at the Earth. }
\label{fig:neutrinoFlare}
\end{figure}

\noindent Figure \ref{fig:neutrinoSteady} shows this quantity, weighted by the squared energy, during the plateau phase ($\sim 15$ days), and Fig. \ref{fig:neutrinoFlare} shows the evolution of the neutrino flux in a flare event.

\section{Detectability of GRO J0422+32 in neutrinos}\label{icecube}

In this section, we study whether the neutrino emission attributed to GRO J0422+42 by the corona model is detectable by the current high-energy neutrino experiments. Our detectability study takes the expected background of atmospheric neutrinos into account at the location of the source, as well as instrumental effects such as detection rate and angular resolution.

\subsection{Instrumental effects}

The small neutrino cross section and the atmospheric neutrino background level are the most important limitations in the discovery of extraterrestrial neutrino signals. For steady sources, the typical period for data analysis in neutrino telescopes is about one year. Here, the observations are background limited, and the sensitivity increases as a function of the exposure time, which is usually added on scales of years. This is not the case for transient sources, where the analyzed period should be on the same timescale as the duration of the burst. If the integration time is too long with respect to the source emitting span, the source disappears entirely in the background of atmospheric neutrinos. On the other hand, integration times that are too short would yield a non significant result because, in this case, the observation is limited by the detector efficiency. Many neutrinos do not interact near the detector, or the events do not pass event selection cuts up to the final neutrino sample. The detector's point spread function (PSF) also has an impact on the detectable fluxes. The effect of the PSF will be to smear the signal from the source over some area, hence lessen the signal to noise ratio. 

\subsubsection{Detection rate}

Our calculations for the detection rate make use of the $\nu_{\mu} + \bar{\nu_{\mu}}$ effective area of the IceCube Neutrino Telescope in its 79-string configuration \citep{odrowski2012}. At the declination of GRO J0422+32, the effective area is conveniently described with the following expression:

\begin{equation}
  \log(A_{\rm{eff}}) = -8.2 + 3.3 \log(E) - 0.25\log(E)^2~\rm{m}^2
\end{equation}

\noindent where $E$ is the neutrino energy in GeV. High-energy neutrino events have a higher detection probability, due to both the proportionality between the neutrino cross section with energy and the long path of the resulting muon, which can reach the detector even if produced outside, thereby increasing the effective volume. 

\begin{figure}
\centering
\includegraphics[clip,width=0.52\textwidth, keepaspectratio]{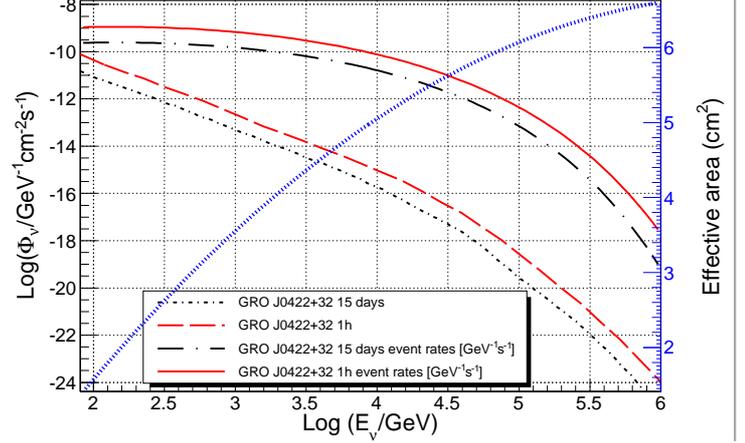}
\caption{Differential flux of neutrinos produced during the 15-day hard state and during the 1-hour flare plotted against the effective area of the IceCube detector. The resulting energy distribution of event rates is also shown.}
\label{fig:RatesAndAeff}
\end{figure}

Figure~\ref{fig:RatesAndAeff} shows the effective area superimposed on the neutrino spectra of GRO J0422+32 from the 1-hour flare and from the 15-day plateau. Although IceCube reaches energies down to 10 GeV \citep{wiebusch2009}, we use a low-energy threshold of 100 GeV. Neutrino energies below this threshold are not suitable for neutrino astronomy because of the large neutrino-muon vertex angle. The expected event rates as a function of the energy, obtained after the multiplication of the two functions, are also represented in this figure. Above 300 GeV, the total neutrino event rate is $3.62 \times 10^{-6}$ Hz, for the 1h flare, and $7.72 \times 10^{-7}$ Hz during the 15-day plateau. As a consequence of the energy cutoff at $\sim$ 8 TeV, neutrino efficiencies for temporal scales of hours, even days, are too low, and if the corona model is valid, the source would not be detected. 

In what follows, we examine the minimum flux required to discover a neutrino signal as a function of the integration time.

\subsection{Atmospheric neutrino background and discovery flux}

To claim a detection, the number of events observed from the location of GRO J0422+32 must overcome the background fluctuations at a 5$\sigma$ level, meaning that the probability of observing this many events in a pure background-only case is below $2.8\times 10^{-7}$. Atmospheric neutrinos do not show significant seasonal variations ($<$1\%) \citep{IceCubeSeasonalAtmo2005}, so we use a Poisson model with a constant rate through the assumed period of data taken for the statistical description of the background. We use the statistics of event counts within a circular area around the source in order to set the 5$\sigma$ level as a function of time. Then, the probability of finding $\emph{k}$ background events within the search area $\Omega$ is the Poisson probability of getting $\emph{k}$ events, given the mean rate of atmospheric neutrinos at the declination of GRO J0422+32.

Atmospheric neutrinos do not arrive at the detector isotropically, but with an energy and zenith dependent component, because of both the direction of the cosmic ray shower through the atmosphere, and, at $E>10$ TeV, attenuation of the neutrino flux by the Earth. The rate of atmospheric neutrino events within an area $\Omega$ around the location of GRO J0422+32 is then estimated with

\begin{equation}
  R_{bg} =  \int_{\Omega}{d\Omega}\int \Phi_{bg}(E,\theta)A_{\rm{eff}}(E,\theta) dE     
\label{eq:bgrates}                       
\end{equation}
where $\Phi_{bg}(E,\theta)$ is the atmospheric neutrino spectrum at declination $\theta$, and $A_{eff}$ is the detector's effective area (equation 21). 

The atmospheric neutrino energy spectrum for $\nu_{\mu}$ ($\bar{\nu_{\mu}}$), averaged over the declination range $7^{\circ} - 90^{\circ}$, has been measured by the IceCube neutrino telescope in the energy range between 100 GeV to 400 TeV \citep{IceCubeAtmo2011}. GRO J0422+32 is located at $\theta = 32^{\circ}$, where the energy distribution of atmospheric neutrinos is represented well by the zenith-averaged spectrum.  

The area over which we perform the integration of equation \ref{eq:bgrates} is the one that minimizes the required source strength for point-source emission, after assuming a PSF.  In the case of IceCube in the 79-string configuration, the PSF for sources with a soft spectrum or an energy cutoff below 50 TeV has a median of $\sim 1^{\circ}$ \citep{odrowski2012}. We assume this PSF with a Gaussian profile. The optimal search area we find has a radius of $1.75^{\circ}$, corresponding to a $\sim$80\% signal retention. The rate of atmospheric neutrinos at the location of GRO J0422+32 within this area is $R_{bg} = 1.06 \times 10^{-6}$ Hz. 

\begin{figure}
\centering
\includegraphics[clip,width=0.52\textwidth, keepaspectratio]{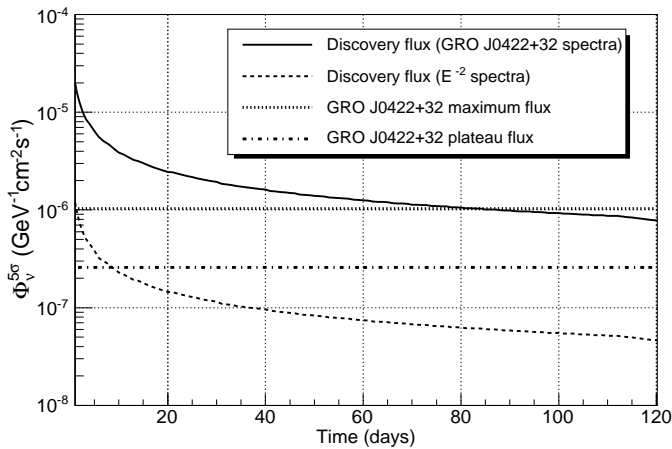}
\caption{Discovery Flux $\Phi_0$, where $\Phi_{\nu} = \Phi_0E^{-2.16}\rm{exp}(-(E/8\rm{TeV})^{-0.52})$ for the case of GRO J0422+32 and an unbroken spectrum $\Phi_{\nu} = \Phi_0E^{-2}$ used for comparison.}
\label{fig:DiscoveryFlux}
\end{figure}

The discovery flux is calculated as the minimum flux required to produce, in 50\% of synthetic data sets, a 5$\sigma$ excess of events within the search area $\Omega$ around GRO J0422+32, during the integration time $\Delta t$. In each of the simulated data sets the number of background events within $\Omega$ is drawn from a Poisson distribution with mean equal to the product of $R_{bg}\times\Delta t$, and signal events are injected in $\Omega$ following the assumed Gaussian PSF with $\sim 1^{\circ}$ median. Figure \ref{fig:DiscoveryFlux} shows the discovery flux as function of the integration time $\Delta t$ on scales of days for two source spectra: the one from GRO J0422+32 in the corona model, following E$^{-2}$ with an energy cutoff at 8 TeV, and an unbroken E$^{-2}$ spectrum for comparison. The effect of the energy cutoff in the detectability of the source is evident in this figure. Both the strongest neutrino flare of 1h duration and the source during the $\sim$20 day hard-state remain undetected in the $\sim$8 TeV cutoff emission scenario. However, if somehow the same or even a smaller amount of energy goes into the acceleration of less relativistic particles but reaching higher energies, close to PeV, the source could be detected.  

\section{Discussion}

Under the physical conditions adopted in our model, the main result obtained is that, if an outburst with similar characteristic to the one observed in 1992 takes place in the present, the probability of detecting it with IceCube is very low. This is due, mainly, to the very short duration of the high-energy flares observed in this source (less than a day).

There are, however, many factors that can enhance the neutrino emission in sources like GRO J0422+32. For example, if the acceleration efficiency increases, then relativistic protons will be able to achieve higher energies, where the neutrino cross-section increases. This will be possible with higher values of the magnetic field in the corona.

Another possibility may be a larger hadron content in the plasma, since neutrinos are produced in hadronic interactions. However, in our model we are already considering $a=100$, which is the ratio observed in cosmic rays, so there is no physical reason for taking $a > 100$. 

Since the main restriction in our study for detectability was the short period of time of activity, is it straightforward to consider longer outbursts. If the neutrino flux remains as in the peak of the event (Fig. \ref{fig:neutrinoFlare}, $t=1$ h) along $\sim 80$ days (see Fig. \ref{fig:DiscoveryFlux}), IceCube will in fact be capable of detecting the source. It will also be detectable if it remains longer in the plateau phase, $\sim 690$ days.

This is the most likely scenario. There are several systems like GRO J0422+32 that shows high-energy episodes with the characteristics needed to enhance the neutrino emission. 

A clear example is GRO J1719-24, which is also a low-mass X-ray binary system that displayed similar gamma-ray spectral characteristics to those observed in Cygnus X-1 and GRO J0422+32 \citep{ling2005}. This source was detected in 1993, during an X-ray  outburst that lasted $\sim 1000$ days with a plateau phase of $\sim 80$ days \citep{ballet1993} . Though the flares observed were also short, they were numerous, so the contribution of all flares may be significant.  

Other examples of this are the low-mass X-ray binaries XTE J1118+480 and GX 339-4. XTE J1118+480 is a transient XRB, which has shown two outburst since its discovery. The first outburst was in 2000 and lasted for about seven months, and the second was in 2005 and lasted for one to two months \citep{vila2012}. GX 339-4 is a well studied system that was detected in the low-hard state on several occasions (1997, 1999 and 2002). The observed X-ray fluxes yield luminosities of up to $L \sim 10^{37}$ erg s$^{-1}$ \citep{vila2010}. The high luminosities detected during the outbursts of both sources and, additionally, the proximity of XTE J1118+480 ($d = 1.72$ kpc), make these systems good Galactic candidates as neutrino sources.

The main difference in these sources is that they present relativistic jets, whereas there is no evidence of a radio jet in GRO J0422+32. Neutrinos might also be produced in the jet if it has a hadron component. Until present, there is no clear evidence to indicate the composition of relativistic jets, although many studies support a hadronic content \citep{heinz2008,romero2008}. Future high-energy detectors, such as CTA, may shed light on this aspect of X-ray binaries.

\section{Conclusions}

We applied a magnetized corona model to describe the spectrum of the low-mass X-ray binary GRO J0422+32 during the outburst of 1992. The presence of nonthermal populations of electron and protons and their interactions with the different fields of the source can explain the high-energy emission detected during the hard state of the source. The model has two main free parameters: the fraction of power injected in relativistic particles ($q$) and the hadron-to-lepton energy ratio ($a$). The best-fit to the spectra is obtained with $a=100$ and $q=0.1$.

Then, we studied the source during a nonthermal flare, produced by an increase in the power injected in relativistic particles. In particular, we considered $q=0.15$, which is the maximum energy available for accelerating particles by magnetic reconnection in our model. The flare duration was taken to be shorter than a day which was limited by the available spectra.

We also estimated the neutrino emission during a flare event, and during the plateau phase of the outburst. Given the short time of activity of the source, a more detailed analysis to study detectability with IceCube is needed.

Although we conclude that the detection of neutrino from an episode like the one studied here is very unlikely, longer events in other Galactic sources may be detectable in the future by IceCube.

\section*{Acknowledgments}

The authors are grateful to James Ling for providing the data of the BATSE instrument. F.L.V thanks Orlando Peres for his insightful comments on neutrino oscillations and the Departament d'Astronomia i Meteorologia, Universitat de Barcelona, where part of this work was carried out. Yolanda Sestayo acknowledges the financial support from J.M. Paredes through ICREA Academia. This work was partially supported by the Argentine Agencies CONICET (PIP 0078) and ANPCyT (PICT 2007-00848), as well as by the Spanish Ministerio de Ciencia e Innovaci\'on (MICINN) under grant AYA2010-21782-C03-01.

\bibliographystyle{aa}  
\bibliography{myrefs2}   

\end{document}